\documentclass[prl,twocolumn,showpacs,amsmath,amssymb]{revtex4}
\usepackage[dvips]{graphicx}
\usepackage{enumerate}
\newcommand \beq{\begin{eqnarray}}
\newcommand \eeq{\end{eqnarray}}

\def\simge{\mathrel{%
       \rlap{\raise 0.511ex \hbox{$>$}}{\lower 0.511ex \hbox{$\sim$}}}}
\def\simle{\mathrel{
       \rlap{\raise 0.511ex \hbox{$<$}}{\lower 0.511ex \hbox{$\sim$}}}}

\begin{document}

\title{ Melting Pattern of 
Diquark Condensates in Quark Matter}
\author{K. Iida,$^1$  T. Matsuura,$^2$  
 M. Tachibana,$^1$ and T. Hatsuda$^2$}
\affiliation{$^{1}$
The Institute of Physical and Chemical Research (RIKEN),  
 Wako, Saitama 351-0198, Japan\\
$^{2}$Department of Physics, University of Tokyo,
  Tokyo 113-0033, Japan}

\begin{abstract}
     Thermal color superconducting phase transitions in high density 
three-flavor quark matter are investigated in the Ginzburg-Landau 
approach.  Effects of nonzero strange quark mass, electric and color charge 
neutrality, and direct instantons are considered.  Weak coupling calculations 
show that an interplay between the mass and electric neutrality effects near 
the critical temperature gives rise to three successive second-order phase 
transitions as the temperature increases: a modified color-flavor locked 
(mCFL) phase ($ud$, $ds$, and $us$ pairings) $\to$ a ``dSC'' phase ($ud$ and 
$ds$ pairings) $\to$ an isoscalar pairing phase ($ud$ pairing) $\to$ a normal 
phase (no pairing).  The dSC phase is novel in the sense that while all eight 
gluons are Meissner screened as in the mCFL phase, three out of nine quark 
quasiparticles are gapless.
\end{abstract}
\maketitle

    Unraveling the phase structure at high baryon density is one of the most 
challenging problems in quantum chromodynamics (QCD).  Among others, 
color superconductivity in cold dense quark matter has been discussed from 
various viewpoints \cite{BL,RWA}.  In relation to real systems such as newly 
born compact stars in stellar collapse, it is important to study the
color superconductivity not only as a function of the quark chemical potential
$\mu$ but also as a function of the temperature $T$.  This is because the 
possible presence of color superconducting quark matter in a star affects the
star's thermal evolution \cite{RST}.
  
    The purpose of this Letter is to investigate phase transitions in
color superconducting quark matter with three flavors
($uds$) and three colors ($RGB$) near the transition temperatures. 
We consider a realistic situation in which nonzero strange quark mass $m_s$, 
electric and color charge neutrality, and direct instantons take effect.  
As we shall see in weak coupling ($m_s, 
\Lambda_{\rm QCD} \ll \mu$), the effects of nonzero $m_s$ and electric
neutrality are important in that they induce multiple 
phase transitions that change the pattern of diquark pairing as $T$ increases.  
In particular, we find a new phase, which we call ``dSC," as an interface between
a modified type of color-flavor locked (mCFL) phase and an isoscalar 
two-flavor (2SC) phase.  

     Throughout this Letter, we adopt the Ginzburg-Landau 
(GL) approach near the transition temperatures, which was previously used to 
study the massless three-flavor case \cite{IB,IB3,MIHB} and is 
a more advantageous framework to weak coupling calculations than 
other mean-field approaches \cite{SRP,abuki}.  In a realistic situation,
the GL potential acquires the following corrections.  First of 
all, nonzero $m_s$ affects the potential through the $s$ quark propagator \cite{abuki}
in such a way as to lower the temperature at which a diquark condensate with 
$s$ quarks dissolves.  This is because the pairing interaction
due to one-gluon exchange is effectively diminished by $m_s$ if 
the pair contains the $s$ quark.  Secondly, when quark matter with 
nonzero $m_s$ is beta equilibrated and neutralized by electrons near the 
transition temperatures, the chemical potentials between $d,s$ quarks 
and $u$ quarks differ.  
Through this chemical potential difference, the GL potential 
acquires another $m_s$ dependence, which is the essential origin of the dSC phase.  
Thirdly, the instanton contribution gives an $m_s$ dependence to the GL 
potential through the effective four-fermion interaction proportional to 
$m_s$.  We finally note that in weak coupling, color neutrality makes  
negligible difference in the GL potential near the transition temperatures \cite{IB}.
 
       We assume that diquark pairing takes place 
in the color-flavor antisymmetric channel with $J^P=0^+$, which is predicted to
be the most attractive channel in weak coupling \cite{brown}.  In this case,
the pairing gap of a quark of color $b$ and flavor $j$ with that of color $c$ 
and flavor $k$ has the form
 $\phi_{bcjk}=\epsilon_{abc} \epsilon_{ijk} ({\mathbf{d}_a})^i$ \cite{IB}.
Here the 3$\times $ 3 matrix $({\mathbf{d}_a})^i$ transforms as a vector 
under $G = SU(3)_C \times SU(3)_{L+R}\times U(1)_B$ and belongs to the 
($3^*,3^*$) representation of $SU(3)_C  \times SU(3)_{L+R}$.
 
       For a homogeneous system of massless quarks $(m_{u,d,s}=0)$, 
the GL potential is invariant under $SU(3)_C \times SU(3)_{L+R}$ and valid
near the critical temperature, $T_c$, common to all states belonging to
the channel considered here. This potential reads \cite{IB,PIS00}
\begin{eqnarray}
\label{GL}
   S= \bar{\alpha} \sum_{a}|{\mathbf{d}_a}|^2
   +\beta_1(\sum_{a}|{\mathbf{d}_a}|^2)^2
   +\beta_2\sum_{ab}|{\mathbf{d}_a}^{\ast}
    \cdot {\mathbf{d}}_b|^2,
\end{eqnarray}
where $({\mathbf{d}_a})^i=(d_a^u, d_a^d, d_a^s)$ and
the inner product is taken for flavor indices.  In the weak coupling regime, 
the coefficients are \cite{IB}
\begin{eqnarray}
\label{b1b2}
 \beta_{1}=\beta_{2}
  =\frac{7\zeta(3)}{8(\pi T_c)^{2}}N(\mu)
 \equiv \beta, \ \
  \bar{\alpha}=4 N(\mu) t \equiv \alpha_0  t, 
\end{eqnarray}
where $N(\mu) = \mu^2/2\pi^2$ is the density of states at the Fermi surface, and
$t=(T-T_c)/T_c$ is the reduced temperature.  With the parameters 
\ (\ref{b1b2}), one finds a second order phase transition at $T=T_c$ from 
the CFL phase ($ {d}_a^i \propto \delta_a^i$) to the 
normal phase ($ {d}_a^i = 0$) in mean-field theory \cite{IB}.  
  
    Let us now consider the effect of a nonzero $m_s$ in the quark propagator on
the GL potential.  We assume $m_{u,d}=0$ for simplicity and consider the high 
density regime, $m_s \ll \mu$.  Near $T_c$ the leading effect of $m_s$ is to 
modify the quadratic term in the GL potential (\ref{GL}).  The corrections to 
the quartic terms are subleading and negligible.  Since $m_s$ affects only
$us$ and $ds$ pairings, the correction to the quadratic term
has the form
\begin{eqnarray}\label{epsilon}
\epsilon 
\sum_{a} ( |d_a^u|^2 + |d_a^d|^2 )
= \epsilon \sum_{a} (|{\mathbf{d}_a}|^2 -|d_a^s|^2).   
\end{eqnarray}
Note that this correction
induces an asymmetry in the flavor 
structure of the CFL phase.
 
     In weak coupling, $\epsilon$ can be calculated by including $m_s$ in the 
Nambu-Gor'kov quark propagator when evaluating the thermodynamic potential.
Following Ref.\ \cite{IB}, we expand the thermodynamic potential not only in 
$d_a^i$ but also in $m_s$ up to ${\cal O}(m_s^2)$, and obtain 
 \begin{eqnarray}
\label{epsilon-wc}
\epsilon  \simeq   
\alpha_0 \frac{m_s^2}{4 \mu^2}
\ln \left(\frac{\mu}{T_c}\right)
\sim 2 \alpha_0  \sigma.
\end{eqnarray}
Here the dimensionless parameter $\sigma$ is given by
\begin{eqnarray}
\sigma = \left( \frac{3 \pi^2} {8 {\sqrt 2}} \right) \frac{m_s^2} {g \mu^2},
\end{eqnarray} 
where $g$ is the QCD coupling constant. As long as $\sigma \ll 1$,
which is relevant at asymptotically high density, the following GL analysis near $T_c$
is valid.  In the latter estimate in Eq.\ (\ref{epsilon-wc}) we use the leading-order 
result in $g$, $\ln (T_c/\mu) \sim -{3\pi^2}/(\sqrt{2} g)$ \cite{brown}.
This behavior of $T_c$ originates from the long-range color magnetic interaction
which prevails in the relativistic regime.  As a result of the modification by $m_s$
to $\ln (T_c/\mu)$, $\epsilon$ has a positive value such that 
$ud$ pairing is favored over $us$ and $ds$ pairings.  Consequently,
the CFL phase becomes asymmetric in flavor space and its critical 
temperature is lowered, leading to the appearance of the 2SC phase 
(${d}_a^i \propto \delta^{is}$) just below $T_c$ 
\cite{abuki}. Note that Eq.\ (\ref{epsilon}) 
has no effect on the 2SC phase. We also note that $T_c$ itself is modified by $m_s$
through the modification of the normal medium as $T_c(1+{\cal O} (g \sigma))$.
  
      We turn to discuss effects of charge neutrality which also depend on 
$m_s$ as mentioned above.  Under beta equilibrium and charge neutrality, the 
electron chemical potential $\mu_e$ and the shift $\delta\mu_i$ of the 
chemical potential of flavor $i$ from the average ($\mu$) are related by
$\delta \mu_i =-q_i\mu_e$, with electric charge $q_i$.  We see from 
Ref.\ \cite{IB} that the thermodynamic-potential correction due to 
$\delta\mu_i$ has the form
   \begin{eqnarray}\label{eta}
\eta~ (\frac{1}{3}  \sum_{a}|{\mathbf{d}_a}|^2
 -\sum_{a} |d_a^u|^2). 
\end{eqnarray}
In weak coupling, where one may regard normal quark matter and electrons
as noninteracting Fermi gases, $\mu_e$ is related to $m_s$ as
$\mu_e=m_s^2/4\mu$.  This estimate is valid in the vicinity of $T_c$ where 
corrections to $\mu_e$ by a finite pairing gap affect only the quartic terms in
the GL potential.  By combining this estimate with the weak coupling 
expression for $\eta/\mu_e$ given in Ref.\ \cite{IB}, we obtain
 \begin{eqnarray}
 \label{eta-wc}
\eta \simeq
 \alpha_0 \frac{m_s^2}{8\mu^2}
\ln \left(\frac{\mu}{T_c}\right)
\sim \alpha_0 \sigma  .
\end{eqnarray} 
Since $\eta>0$, $ds$ pairing is more favorable than $ud$ and $us$ pairings.
This feature stems from the modification by $\delta \mu_i$ 
to the exponential factor of $T_c$, $\exp[-{3\pi^2}/(\sqrt{2} g)]$, 
which tends to increase (decrease) the critical temperature for $ij$
pairing when $\delta\mu_i+\delta\mu_j >0 (<0)$.

    We consider color neutrality of the system as well.  In 
contrast to the case at $T=0$, however, it affects only the quartic terms in 
the GL potential through possible chemical potential differences between 
colors \cite{IB,RIS}, and in weak coupling its magnitude is suppressed by 
${\cal O}((T_c/g\mu)^2)$ compared to the leading quartic terms.  Thus color 
neutrality has no essential consequence to the phase transitions considered in
this Letter.  A major difference between corrections of the charge neutrality and the color 
neutrality is that the former shifts the quark chemical potentials even in the
normal phase, while the latter works only when the pairing occurs.  This is 
why the former is more important than the latter near $T_c$.
   
     The direct instanton at nonzero $m_s$, which induces an effective 
four-fermion interaction between $u$ and $d$ quarks \cite{Sch}, leads to a
quadratic term in the GL potential,
$\zeta~\sum_a |d_a^s|^2$. 
An explicit weak coupling calculation shows that $\zeta\sim -\alpha_0 (m_s/\mu) 
(\Lambda_{\rm QCD}/\mu)^9 (1/g)^{14}$.  The negative sign indicates that the 
instanton favors $ud$ pairing as does one-gluon exchange [see Eq.\ 
(\ref{epsilon})], but the magnitude of $\zeta$ is highly suppressed at high densities. 
Hereafter we will thus ignore instanton effects.

     Since the two effects of nonzero $m_s$, characterized by Eqs.\ 
(\ref{epsilon}) and (\ref{eta}), favor $ud$ pairing and $ds$ pairing, 
respectively, the finite temperature transition from the CFL to 
the normal phase at $m_s=0$ is expected to be significantly modified.
  In fact, successive color-flavor unlocking transitions take 
place instead of a simultaneous unlocking of all color-flavor combinations.
To describe this {\em hierarchical thermal unlocking}, it is convenient to  
introduce a parameterization,
 \begin{eqnarray}
\label{s}
d_a^i
=
 \left(
  \begin{array}{lll} 
      \Delta_1&0&0    \\
  0&\Delta_2&0    \\
  0&0&\Delta_3    \\
  \end{array}
 \right),
  \end{eqnarray}
where $\Delta_{1,2,3}$ are assumed without loss of generality to be real.  We 
also name the phases for later convenience as
 \begin{eqnarray}
\label{phase-def} 
  \begin{array}{llcl} 
     \Delta_{1,2,3} \neq 0 & & :  &  {\rm mCFL}, \\
\Delta_1=0, & \Delta_{2,3}\neq 0   & :& {\rm uSC},      \\
\Delta_2=0, &  \Delta_{1,3} \neq 0 & :& {\rm dSC},      \\
\Delta_{1,2}=0, & \Delta_3 \neq 0  &: & {\rm 2SC},                
  \end{array}
  \end{eqnarray} 
where dSC (uSC) stands for superconductivity in which for $d$ ($u$) quarks all
three colors are involved in the pairing.
   
     In terms of the parameterization (\ref{s}), the GL potential with 
corrections of ${\cal O}(m_s^2)$ to the quadratic term, Eqs.\ (\ref{epsilon}) and 
(\ref{eta}), reads
\begin{eqnarray}
 \label{new-GL}
S&=&\bar{\alpha}' (\Delta_1^2+\Delta_2^2+\Delta_3^2) 
- \epsilon \Delta_3^2 - \eta \Delta_1^2\nonumber \\
 &+& \beta_1 (\Delta_1^2+\Delta_2^2+\Delta_3^2)^2
 + \beta_2 ( \Delta_1^4+\Delta_2^4+\Delta_3^4),
\end{eqnarray}
where 
$\bar{\alpha}' =\bar{\alpha}+\epsilon+\frac{\eta}{3}$.
 
     We proceed to analyse the phase structure dictated by Eq.\ 
(\ref{new-GL}) with the weak coupling parameters (\ref{b1b2}), 
(\ref{epsilon-wc}), and (\ref{eta-wc}) up to leading order in $g$.
 In Figs. 1 and 2 the results obtained by 
solving the coupled algebraic equations, $\partial S/\partial\Delta_{1,2,3}=0$,
are summarized.  Figure 1(a) shows the second-order phase transition, CFL $\to$ 
normal for $m_s=0$.  Figures 1(b,c) represent how the phase transitions and 
their critical temperatures bifurcate as we introduce (b) effects of a nonzero 
$m_s$ in the quark propagator and then (c) effects of charge neutrality.
In case (b), two second-order phase transitions arise, 
mCFL (with $\Delta_1=\Delta_2$) $\to$ 2SC at 
$T=T_c^s \equiv (1-4 \sigma)T_c$, 
and 2SC $\to$ normal at $T=T_c$.  In case (c), there arises three 
successive second-order phase transitions, mCFL $\to$ dSC at $T=T_c^{\rm I}$,
dSC $\to$ 2SC at $T=T_c^{\rm II}$, and 2SC $\to$ normal at $T=T_c^{\rm III}$. 
Shown in Fig.\ 2 is the $T$-dependence of the gaps $\Delta_{1,2,3}$ for 
 case (c).  All the gaps are continuous functions of $T$, but their slopes
are discontinuous at the critical points, which reflects the second order 
nature of the transitions in the mean-field treatment of Eq.\ (\ref{new-GL}).

\begin{figure}[t]
\begin{center}
\includegraphics[width=7cm]{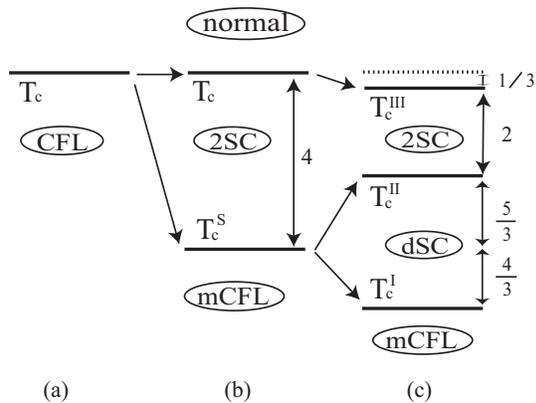}
\end{center}
\vspace{-0.5cm}
\caption{Transition temperatures of the 
three-flavor color superconductor in weak coupling:
(a) all quarks are massless;
(b) nonzero $m_s$ in the quark propagator is considered;
(c) electric charge neutrality is further imposed.
The numbers attached to the arrows are in units of $\sigma T_c$.
}
\label{fig:Tc}
\end{figure}

\begin{figure}[t]
\begin{center}
\includegraphics[width=7cm]{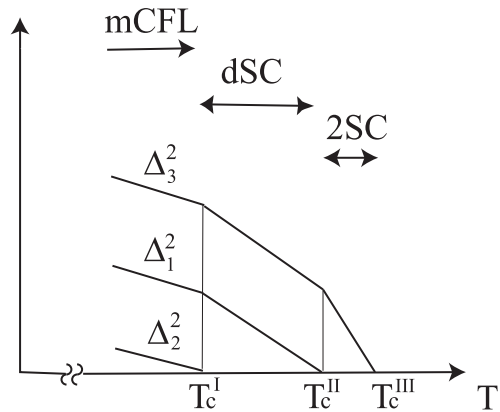}
\end{center}
\vspace{-0.5cm}
\caption{A schematic illustration of the gaps squared
 as a function of $T$. }
\label{fig:conden}
\end{figure}

    We may understand the bifurcation of the transition temperatures in 
weak coupling as follows.  In the massless case (a), $T_c$ is degenerate 
between the CFL and 2SC phases, the chemical potential is common to all
three flavors and colors, and the CFL phase is more favorable than
the 2SC phase below $T_c$.  As one goes from (a) to (b), nonzero $m_s$ sets in, which
tends to suppress the pairing interactions including the $s$ quark.  
The critical temperature for the CFL phase is then lowered, which allows the
2SC phase to appear at temperatures between $T_c^s$ and $T_c$.  As one goes 
from (b) to (c), charge neutrality sets in, which acts to decrease the 
chemical potential of $u$ quarks by $(2/3)\mu_e$ and increase that of 
$d$ and $s$ quarks by $(1/3)\mu_e$.  Since $\mu_e>0$, the average chemical 
potential of quarks involved in $ds$ pairing increases, while those in $ud$ and
$us$ pairings decrease equally.  Accordingly, the transition temperatures
further change from $T_c$ to $T_c^{\rm III}$ and from $T_c^s$ to 
$T_c^{\rm I}$ and $T_c^{\rm II}$. 

     Now we examine in more detail how the color-flavor unlocking in case 
(c) proceeds with increasing $T$ from the region below $T_c^{\rm I}$.

 \noindent
 (i)
     Just below $T_c^{\rm I}$, we have a CFL-like phase, but the three gaps take different 
values, with an order $\Delta_3 > \Delta_1 > \Delta_2 \neq 0$ (the mCFL phase).
The reason why this order is realized can be understood from the GL potential  
(\ref{new-GL}).  The $\epsilon$-term and $\eta$-term in Eq.\ (\ref{new-GL}) 
tend to destabilize $us$ pairing ($\Delta_2$)
relative to $ud$ pairing ($\Delta_3$) and $ds$ pairing ($\Delta_1$).  
Since $\epsilon > \eta (> 0)$, $ds$ 
pairing is destabilized more effectively than $ud$ pairing.  The value of each 
gap in the mCFL phase reads
     \begin{eqnarray}      
\Delta_3^2&=&\frac{\alpha_0}{8\beta} \left( 
 \frac{T_c-T}{T_c} 
+ \frac{8}{3} \sigma 
\right) , \\
\Delta_1^2 &=&\frac{\alpha_0}{8\beta} \left(
 \frac{T_c-T}{T_c} 
-\frac{4}{3}\sigma
\right)  , \\
 \Delta_2^2 &=& \frac{\alpha_0}{8\beta} 
\left(
 \frac{T_c-T}{T_c} 
- \frac{16}{3}\sigma
\right)  .
\end{eqnarray}
The mCFL phase has only a global symmetry $U(1)_{C+L+R} \times U(1)_{C+L+R}$ 
in contrast to the global symmetry $SU(3)_{C+L+R}$ in the CFL phase with 
$m_{u,d,s}=0$.  There are no gapless quark excitations in both mCFL and CFL 
phases.  As $T$ increases, the first unlocking transition, the unlocking of 
$\Delta_2$ (the pairing between $Bu$ and $Rs$ quarks), takes place at the critical temperature,
 \begin{eqnarray}
 \label{Tc-1}
T_c^{\rm I}  = \left( 1 - \frac{16}{3} \sigma \right) T_c  .
\end{eqnarray}

\noindent
(ii) For $T_c^{\rm I} < T < T_c^{\rm II}$, $\Delta_2=0$ and
\begin{eqnarray} 
\Delta_3^2&=& \frac{\alpha_0}{6 \beta} 
\left(
 \frac{T_c-T}{T_c} 
+\frac{2}{ 3}\sigma
\right) , \\
\Delta_1^2&=& \frac{\alpha_0}{6 \beta} 
\left( \frac{T_c-T}{T_c}-\frac{7}{3} \sigma \right)  .
\end{eqnarray}
In this phase, we have only $ud$ and $ds$ pairings (the dSC phase), and there 
is a manifest symmetry, $U(1)_{C+L+R} \times U(1)_{C+L+R} \times 
U(1)_{C+V+B} \times U(1)_{C+V+B}$.  By diagonalizing the inverse quark 
propagator in the Nambu-Gor'kov representation, we find three gapless quark 
excitations in the color-flavor combinations: $Bu$, $Rs$, and a 
linear combination of $Ru$ and $Bs$.  At $T=T_c^{\rm II}$, the second 
unlocking transition, the unlocking of $\Delta_1$ (the pairing between
$Gs$ and $Bd$ quarks), takes 
place at the critical temperature,
   \begin{eqnarray}
 \label{Tc-2}
T_c^{\rm II}  = \left( 1 -\frac{7}{3}\sigma \right) T_c.
\end{eqnarray} 

\noindent
(iii) For $T_c^{\rm II} < T < T_c^{\rm III}$, 
one finds the 2SC phase, which has only $ud$ pairing with 
\begin{eqnarray}
\Delta_3^2=\frac{\alpha_0}{4 \beta} 
\left(
 \frac{T_c-T}{T_c} 
-\frac{1}{3}\sigma
\right) .  
\end{eqnarray} 
The 2SC phase has a symmetry $SU(2)_C \times SU(2)_{L+R} \times U(1)_{C+B} 
\times U(1)_{L+R+B}$ \cite{MIHB}.  In this phase the $s$ quark and $B$ quark 
excitations are gapless.  The final unlocking transition where 
 $\Delta_3$ (the pairing between $Rd$ and $Gu$ quarks) vanishes occurs at 
\begin{eqnarray}
 \label{Tc-3}
T_c^{\rm III}  = \left( 1 -\frac{1}{3}\sigma \right) T_c .
\end{eqnarray}  
Above $T_c^{\rm III}$, the system is in the normal phase.
  
\begin{table}[b]
\caption{The symmetry, the gapless quark modes, and the number of Meissner screened 
gluons in the mCFL, dSC, and 2SC phases.  The gapless quark mode $(Ru,Bs)$ 
denotes the linear combination of $Ru$ and $Bs$ quarks.} 
\begin{tabular}{|c|c|c|c|}
\hline 
  & Symmetry & Gapless  &  Number of  \\
  &   &   quark modes & massive gluons\\
\hline \hline
mCFL &$[U(1)]^2 $ &  none  & 8\\
\hline
dSC   &  $[U(1)]^4 $ & 
$Bu$, $Rs$ & 8\\
&  &  ($Ru,Bs$) &\\
\hline
2SC   & $[SU(2)]^2 \times [U(1)]^2 $
  & $Bu$, $Bd$, $Bs$,&   5        \\
&  & $Rs$, $Gs$&           \\
\hline
\end{tabular}
 \end{table}

     In Table I, we summarize the symmetry and the gapless quark modes in each
phase discussed above.  The number of gluons having nonzero Meissner masses,
which is related to the remaining color symmetry, is also shown \cite{MITH}.  
We note that more gapless quark modes may appear if the system is in the  
close vicinity of $T_c^{\rm I}$, $T_c^{\rm II}$, and $T_c^{\rm III}$ where 
$\Delta_2$, $\Delta_1$, and $\Delta_3$ are less than $\sim m_s^2/\mu$ 
\cite{gapless}.

     So far, we have studied the phase transitions in the mean-field level. 
In weak coupling, as shown in Ref.\ \cite{MIHB} in the massless limit, thermally fluctuating 
gauge fields could change the order of the transitions described in Figs.\ 1 and 2.  A 
detailed account on this effect will be reported elsewhere \cite{MITH}; here we
 recapitulate the important results.  First, the second order transition,
mCFL $\to$ dSC, remains second order even in the presence of the thermal gluon 
fluctuations.  This is because all eight gluons are Meissner screened at $T=T_c^{\rm I}$ 
and thus cannot induce a cubic term with respect to the order 
parameter in the GL potential.  On the other hand, the transitions, dSC $\to$ 
2SC and 2SC $\to$ normal, become weak first order since some gluons, which
are massless in the high temperature phase, become Meissner screened in the low
temperature phase (Table I).
 
     In summary, we have investigated color-flavor unlockings at finite 
temperatures taking into account the strange quark mass and  
charge neutrality in the GL approach.  We find three successive 
unlocking transitions, mCFL $\to$ dSC $\to$ 2SC$\to$ normal, occurring
 in weak coupling.  Most remarkably, the dSC phase appears between the mCFL 
and 2SC phases.  In this phase all eight gluons are Meissner screened and the three quark 
excitations are gapless.  The question of how the phase structure of neutral 
quark matter obtained near $T_c$ is connected to that at $T=0$, which 
approaches the CFL phase in the limit of high density, is an 
interesting open problem.

\begin{acknowledgments}
     We are grateful to G. Baym and H. Abuki for helpful discussions.
This work was supported in part by RIKEN Special Postdoctoral Researchers
Grant No.\ A11-52040 and No.\ A12-52010, and by the Grants-in-Aid of the
Japanese Ministry of Education, Culture, Sports, Science, and Technology
(No.~15540254).
\end{acknowledgments}


\end{document}